 \documentclass[11pt,twocolumn]{article}

 \pdfoutput=1

 \usepackage[english]{babel}
\usepackage{graphicx} 
\usepackage{amsmath}

\usepackage{enumitem}

\usepackage{float}
\usepackage{ifthen}
\usepackage{hyperref}

\usepackage{titlesec}
\titlespacing*{\section}{0pt}{0.2\baselineskip}{\baselineskip}

\begin{document}




\twocolumn[
\begin{center}
{\bf \huge A moist ``available enthalpy'' norm: definition and}
\vspace*{1mm}
\\
{\bf \huge  comparison with existing ``energy'' norms.}\\
\vspace*{3mm}
{\Large \bf by Pascal Marquet and Jean-Fran\c{c}ois Mahfouf} {\Large \it (WGNE Blue-Book 2015)}. \\
\vspace*{2mm}
{\Large M\'et\'eo-France. CNRM/GMAP.
 Toulouse. France.}
{\Large \it E-mail: pascal.marquet@meteo.fr} \\
\vspace*{2mm}
{\large \it Version 2 (\today): an extended version, with units in Fig.1 corrected}
\\
\vspace*{2mm}
\end{center}
]





 \section{\underline{\Large Motivations.}} 
\vspace{-3mm}

Several inner-products, based on energy norms, have been used in preliminary 4-D variational assimilation to minimize cost functions.
It was supposed that the energy corresponding to observational errors could be distributed equally amongst these different basic prognostic fields.

Inner-products based on the same energy norms are also used to define the (dry) semi-implicit operators and the (dry) normal modes of GCMs or NWP models, as far as they are invariant by the linear set of primitive equations.
The same inner-products are currently used for computing dry or moist singular vectors and for determining forecast errors or sensitivity to observations based on tangent linear and adjoint models.

However, it is shown in a paper to be submitted (Marquet and Mahfouf, QJRMS) that these norms suffer from a lack of reliability, since their definitions are not unique and because these ``Total Energy Norms'' are not based on an ``Energy'' concept.

This can be illustrated by considering the sum of temperature and water vapour contributions of the total energy norm defined in Ehrendorfer \textit{et al.\/} (1999, E99) in terms of the quadratic perturbations $(T')^2/2$ and $(q'_v)^2/2$, leading to
\vspace{-0.15cm}
\begin{equation}
      N_{\rm E99}  \: =  \:  \frac{c_{pd}}{T_r} \: 
             \overline{  \frac{(T')^2}{2}  }
     \; + \;
              \: \epsilon(z) \;
              \frac{{(L_{\rm vap})}^2 }{c_{pd}\:T_r} \:
             \overline{    \frac{{(q'_v)}^2}{2}  }
    \; . 
    \label{def_E99}
\end{equation}

The first term (temperature contribution) cannot be derived from the enthalpy which roughly varies as $c_{pd}\: T$ and is linear in temperature.
In fact the formulation retained in (\ref{def_E99}) is based on the Available Potential Energy (APE) of Lorenz, provided that the reference temperature is a constant value $T_r$.
What is usually called ``Total Energy Norm'' should thus be called ``Total APE Norm''.

As for the water-vapour contribution (second term) of the norm (\ref{def_E99}), it is derived from the temperature contribution with the additional hypothesis that changes of temperature and moisture are related by $c_{pd}\:T' \approx - L_{\rm vap}\:q'_v$, namely by assuming conservation of perturbed moist static energy ($c_{pd}\:T \: + \:  L_{\rm vap}\:q_v$) in condensation processes.
This assumption is not realistic, in particular in all under-saturated regions.
For this reason, an arbitrary factor $\epsilon(z)$ is often introduced, which may vary with altitude, though without clear justification.

An alternative definition for the water-vapour contribution of the norm is suggested in Mahfouf and Bilodeau (2007, MB07) by assuming zero departure (at constant pressure) in the relative humidity defined by $q_v / q_{sw}(T,p)$, this implying $q'_v \approx \: ( \Gamma_q ) \: T'$ where 
$\Gamma_q = (\overline{q_v}/\overline{q_{sw}}) \: (\partial \, \overline{q_{sw}} / \partial \, T)$,  leading to
\vspace{-0.2cm}
\begin{equation}
       N_{\rm MB07}  \: =  \:  \frac{c_{pd}}{T_r} \: 
             \overline{  \frac{(T')^2}{2}  }
     \; + \;
       \: \frac{c_{pd}}{T_r}
       \: \frac{1}{(\Gamma_q)^2}
       \:  \overline{    \frac{{(q'_v)}^2}{2} }
    \; . 
          \label{def_Nqv_MB07}
\end{equation}
This definition in terms of $\Gamma_q$ is also arbitrary and may not be valid everywhere.

 \section{\underline{\Large The available enthalpy norm.}} 
\vspace{-3mm}

The dry-air available enthalpy is an Exergy function derived in general thermodynamics.
It is defined as $a_h = (h - h_r) - T_r \: (s-s_r)$ in terms of enthalpy $h$,  entropy $s$ and some reference values ($h_r$, $s_r$) depending on ($T_r$, $p_r$).
It is shown in Marquet (1991, 2003) that integral of $a_h$ can replace the APE of Lorenz, leading to a modified temperature contribution of the norm (\ref{def_E99}) which can be written as
\vspace{-0.1cm}
\begin{equation}
      N_T  
     \: =  \:  
              \frac{c_{pd} \: T_r}{(\: \overline{T} \:)^2} \:
             \overline{  \frac{(T')^2}{2}  }
     \: =  \:  
             \frac{c_{pd}}{T_r} \: 
         \left(  
             \frac{T_r}{\; \overline{T} \;}
         \right)^{\! \! 2} \;
             \overline{  \frac{(T')^2}{2}  }
     \: =  \:  
             \overline{  \frac{(T')^2}{V_T}  }
    \; ,
    \label{def_NT}
\end{equation}
\vspace*{-3mm}
\noindent where the variance of temperature is thus weighted by
\begin{equation}
V_T = \frac{2 \; T_r}{c_{pd}} \:
      \left( \frac{\overline{T}(z)}{T_r} \right)^2 \: .
\label{def_VT}
\end{equation}
The comparison with (\ref{def_E99}) or (\ref{def_Nqv_MB07}), 
where 
\begin{equation}
V_{T/\rm E99} = V_{T/\rm MB07} = \frac{2 \; T_r}{c_{pd}} \: , 
\end{equation}
shows that, since the new additional factor $(\overline{T}/T_r)^2$ is not a constant, $V_T$ may vary with pressure or latitude, via the isobaric (or zonal) mean value $\overline{T}(\varphi, p)$.

A moist version of $a_h$ is defined in Marquet (1993, M93) and it is possible to define a ``Moist Available Enthalpy Norm'' (Marquet, 2005, unpublished results).
The temperature contribution is still given by (\ref{def_NT}) and the new water-vapour contribution is expressed in terms of the mixing ratio, leading to
\vspace{-1.mm}
\begin{equation}
   N_v 
    \; = \; 
             \frac{R_v \: T_r}{\overline{r_v} } \: 
             \overline{  \frac{{(r'_v)}^2}{2} }
    \; = \; 
              \overline{  \frac{{(r'_v)}^2}{V_q} }
      \label{def_Nv}  \; ,
\end{equation}

\vspace*{-3mm}
\noindent where the variance of water is weighted by 
\begin{equation}
V_q = \frac{2 \; \overline{r_v}(z)}{R_v \; T_r} \: .
\label{def_Vv}
\end{equation}
Differently from the constant weighting factor 
\begin{equation}
V_{q/\rm E99} = \frac{2\: c_{pd} \: T_r}{(L_{\rm vap})^2}
\end{equation}
in (\ref{def_E99}), but similarly to the varying one 
\begin{equation}
V_{q/\rm MB07} = \frac{2 \: [\: \Gamma_q(z) \:]^2 \: T_r}{c_{pd}}
\end{equation}
in (\ref{def_Nqv_MB07}),
$V_q$ must clearly decrease with height in (\ref{def_Vv}), due to the isobaric (or zonal) mean value $\overline{r_v}(\varphi, p)$.

 \section{\underline{\Large A numerical study.}} 
\vspace{-3mm}

The datasets from MB07 have been used for comparing the quantities $\sqrt{\:V_T\:}$ and $\sqrt{\:V_q\:}$  estimated from the definition of moist ``energy'' norms (E99, MB07) and present Exergy (Available Enthalpy) norm against standard deviation of analysis increments ($S_T$, $S_q$) derived in MB07 from the CMC 3DVAR system.
Constants as are equal to
\vspace{-3mm}
\begin{tabbing}
 ------ \= -------------------------- \=  ------------------------------\= \kill
\> $T_r\approx 273$~K, \> $c_{pd} \approx 1005$~J/K/kg,   \\
\> $R_v = 462$~J/K/kg, \> $L_{\rm vap} \approx 2.5 \: 10^6$~J/kg.
\end{tabbing}

\begin{figure}[hbt]
\centering
\includegraphics[width=0.75\linewidth]{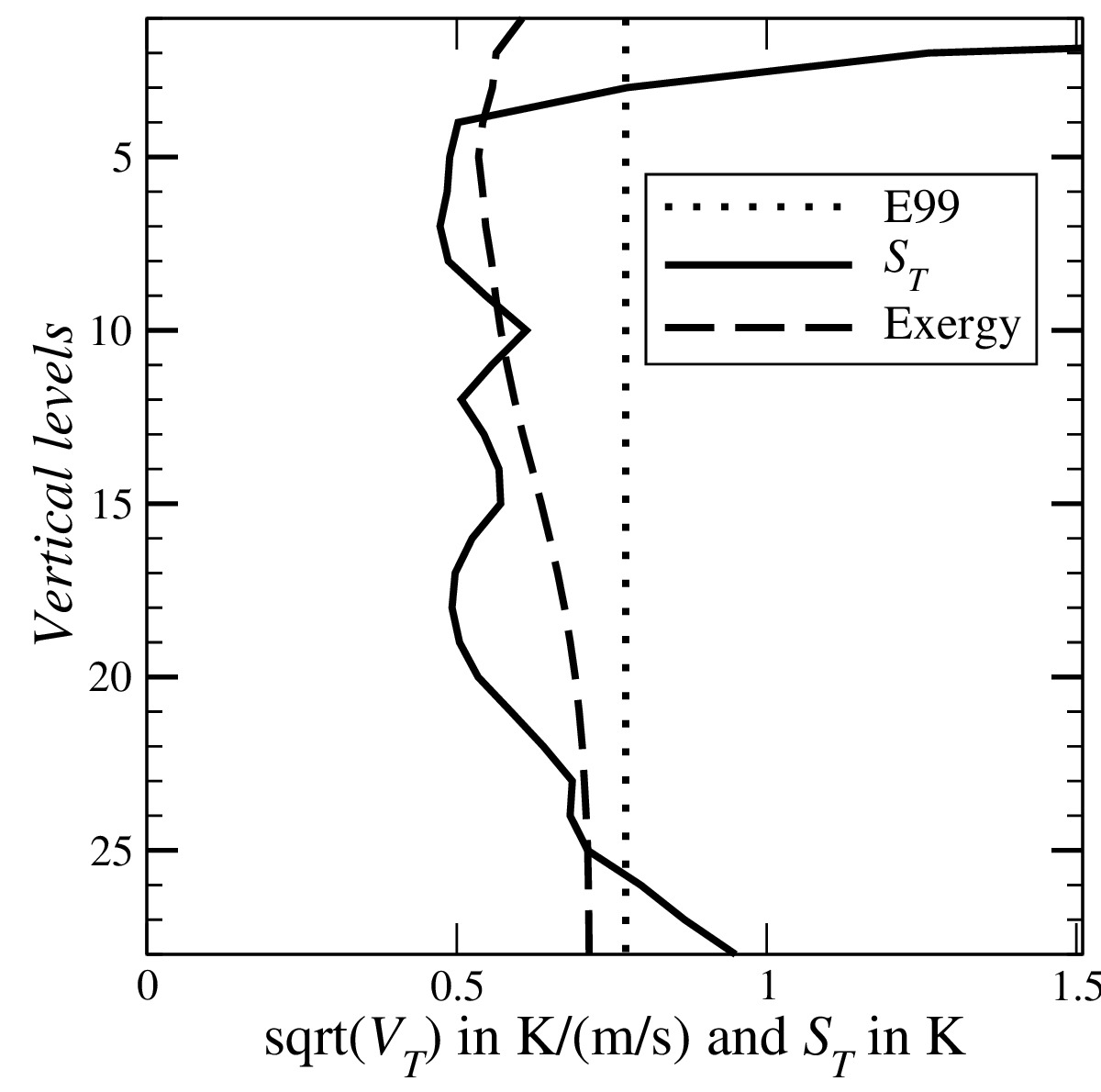}
\includegraphics[width=0.75\linewidth]{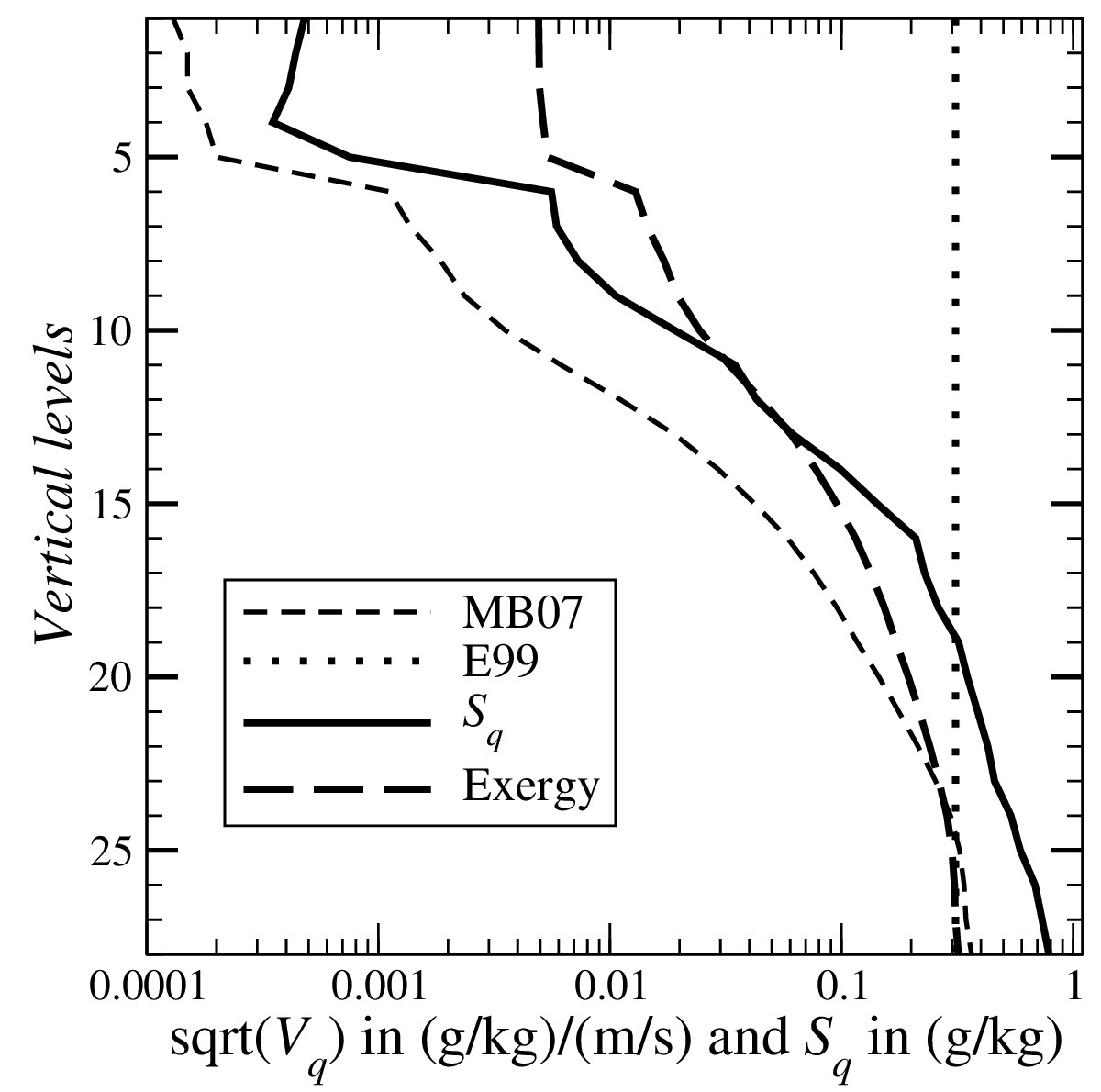}
\vspace{-1mm}
\caption{\small \it Comparison of quantities $\sqrt{\:V\:}$ estimated from the definition of moist norms against standard deviation of analysis increments $S_T$ and $S_q$.
\label{fig_MB07_sigma}}
\end{figure}

It is shown in Fig.\ref{fig_MB07_sigma} (upper panel) that the Exergy norm (\ref{def_NT}) generating a term $\sqrt{\:V_T\:}$ (heavy dashed) varies with height and is in better agreement with $S_T$ (solid) than the constant value $\sqrt{\:V_{T/\rm E99}\:}$ (dotted), in particular within the troposphere (below level $5$).

Similarly, it is shown in Fig.\ref{fig_MB07_sigma} (lower panel) that the Exergy norm (\ref{def_Nv}) generates a term $\sqrt{\:V_q\:}$ (heavy dashed) which varies with height and is in much better agreement with $S_q$ (solid) than the constant value $\sqrt{\:V_{q/\rm E99}\:}$ (dotted), in particular in the troposphere below level $5$.
The norm (\ref{def_Nqv_MB07}) defined in MB07 corresponds to $\sqrt{\:V_{q/\rm MB07}\:}$, which is plotted as thin dashed line.
It is in better agreement with $S_q$, in particular in the stratosphere above level $5$.

 \section{\underline{\Large Conclusions.}} 
\vspace{-5mm}

It has been shown that it is possible to define a moist-air norm based on the concept of ``Available Enthalpy'' (or ``Availability function'' or ``Exergy'').

This new Exergy norm is in close agreement with standard deviation of analysis increments, and a water-vapour contribution is obtained, without the need of unclear and arbitrary assumptions.

The improvement observed with the use of $\sqrt{\:V_T\:}$ in the temperature contribution of the norm (\ref{def_NT}) is a pleasant surprise.
As for the water contribution of the norm (\ref{def_Nv}), the use of $\sqrt{\:V_q\:}$ can explain the observed strong decrease of the norm with height by about two order of magnitude, and this result is obtained without arbitrary assumptions or use of $\epsilon(z)$.

A striking feature is that liquid water or ice cloud contents do not contribute to an independent quadratic norm depending on $(r'_l)^2$ or $(r'_i)^2$.
Only $(r'_v)^2$ must be considered, with $q_l$ and $q_i$ only impacting the moist definition of $c_p$ in factor of the temperature contribution of the norm, with small impacts on the norm itself.

Moreover, the surface pressure contribution which is included in all norm definitions also results from the Available Enthalpy norm (not shown), leading to a complete explanation of all three temperature, surface pressure and water vapour contributions of the new moist-air norm.

The above results justify the proposed method, that is to start with an availability function (which generalizes the APE of Lorenz to the case of a moist atmosphere) in order to build a physically sound moist norm.

\vspace{1mm}
\noindent{\large\bf \underline {References}}
\vspace{0mm}

\noindent{$\bullet$ Mahfouf, J.-F., Bilodeau, B.} {(2007)}.
Adjoint sensitivity of surface precipitation to initial conditions.
{\it Mon. Wea. Rev.}
{\bf 135}:
p.2879--2896. 
\vspace{-0.5mm}

\noindent{$\bullet$ Ehrendorfer,~M., Errico,~R.~M and Reader,~K.,~D.} {(1999)}.
Singular-Vector perturbation growth in a 
primitive equation model with moist physics.
{\it J. Atmos. Sci.}
{\bf 56}:
p.1627--1648.
\vspace{-0.5mm}

\noindent{$\bullet$ Marquet~P.} {(1991)}.
On the concept of exergy and available
enthalpy: application to atmospheric energetics.
{\it Q. J. R. Meteorol. Soc.}
{\bf 117}:
p.449--475.
\vspace{-0.5mm}

\noindent{$\bullet$ Marquet~P.} {(1993)}.
Exergy in meteorology: definition and properties
of moist available enthalpy.
{\it Q. J. R. Meteorol. Soc.}
{\bf 119}:
p.567--590.
\vspace{-0.5mm}

\noindent{$\bullet$ Marquet,~P.} {(2003)}.
The available-enthalpy cycle. I: 
Introduction and basic equations.
{\it Q. J. R. Meteorol. Soc.}
{\bf 129}:
p.2445--2466.


\end{document}